\documentstyle[prl,aps,epsf]{revtex}
\def\myfigure#1#2{{\leftskip=0.000753\textwidth
\rightskip\leftskip\small \begin{figure}\baselineskip=14pt plus 2pt
minus 1pt \centerline{#1}\nobreak\smallskip\nobreak #2\end{figure}}}
\global\firstfigfalse

\begin{document}
\draft 
\twocolumn[\hsize\textwidth\columnwidth\hsize\csname @twocolumnfalse\endcsname
\title {Shape Instabilities in the Dynamics of a Two-component Fluid
Membrane
}
\author{P.\ B.\ Sunil Kumar and Madan Rao}

\address{Institute of Mathematical Sciences, Taramani, Chennai (Madras)
600 113, India}

\date{\today}

\maketitle

\begin{abstract} We study the shape dynamics of a two-component fluid
membrane, using a dynamical triangulation monte carlo simulation and a
Langevin description.  Phase separation induces morphology changes
depending on the lateral mobility of the lipids. When the mobility is
large, the familiar labyrinthine spinodal pattern is linearly unstable
to undulation fluctuations and breaks up into buds, which move towards
each other and merge. For low mobilities, the membrane responds
elastically at short times, preferring to buckle locally, resulting in a
crinkled surface.  
\end{abstract}

\pacs{PACS: 87.22.Bt, 64.60.Cn}
]

A mixture of phospholipid molecules, e.g. DMPC + SOPC, aggregate in an
aqueous solvent to form closed bilayered fluid membranes of dimension $10-20
\,\mu$m. Such mixed vesicles exhibit definite equilibrium shapes depending on
the ambient temperature, osmotic pressure, relative concentration and history
of preparation. Amongst the multitude of equilibrium shape transitions
predicted by theory \cite{JULICHER,TWOCOMP} and observed in phase contrast
video micrography, the most dramatic is the phase segregation induced budding
\cite{JULICHER} at which a large `parent' vesicle sprouts out a spherical bud
consisting of one of the lipid species, attached to it by a narrow umbilical.
Such equilibrium budding has been oberved in a mixture of natural brain
(sphingomyelin) lipids \cite{DOBRY}. 

Apart from these equilibrium studies, there has been relatively little work
on the dynamics \cite{CAI} of mixed fluid membranes \cite{FIRST,TANI}. In the
context of a 2-component fluid membrane undergoing phase separation, the
study of nonequilibrium shapes becomes essential, since the lateral diffusion
coefficient of lipids $10^{-8} - 10^{-7}$cm$^2$/sec, are such that the time
scale over which local lipid concentration relaxes is comparable to shape
relaxation times. In this Letter we study the shape dynamics of a
2-component, open fluid membrane experiencing phase separation, using both a
dynamical triangulation monte carlo simulation (DTMC) and a Langevin
description. 

The shape dynamics of a 2-component fluid membrane involves 6 slow variables,
the conserved total lipid density $\rho = \rho_A+\rho_B$, the conserved
relative concentration $\phi = (\rho _A - \rho _B)/\rho$, the `broken
symmetry' shape variable ${\bf R}(u_1,u_2)$, and the total momentum density
${\vec {\pi}}$, which involves both lipid and solvent hydrodynamics. The
primary dissipation mechanisms arise from the in-plane lipid viscosity and
out-of-plane solvent viscosity \cite{NOTE1}. We include the effects of
solvent hydrodynamics in our Langevin analysis, but not in our DTMC. 

Over length scales larger than the bilayer thickness ($d \sim 40$\AA), a
2-component fluid membrane can be described by a continuum hamiltonian
\cite{JULICHER,TWOCOMP} ${\cal H} = {\cal H}_c + {\cal H}_{c-\phi} + {\cal
H}_{\phi} + {\cal H}_{\rho}$, where the curvature energy \cite{MEMBRANE},
written in terms of the extrinsic curvature $H$ reads ($g \equiv det
(g_{ij})$, where $g_{ij}$ is the metric) 

\begin{equation} 
{\cal H}_c + {\cal
H}_{c-\phi} = \,\,\kappa_c \,\int (H-H_0(\phi))^2 \sqrt g \,d^2u\,\,,
\label{eq:helfrich} 
\end{equation} 

while ${\cal H}_{\phi}$ takes the usual Landau-Ginzburg form,

\begin{equation} 
{\cal H}_\phi = \int \left[ \frac
{\sigma}{2} (\nabla \phi)^2 - \frac {\phi^2}{2} + \frac {\phi^4}{4} \right]
\sqrt g d^2u\,\,. \label{eq:landau} \end{equation} The spontaneous curvature
$H_0(\phi) = c_0 (1 + \phi)/2$, reflects the shape asymmetry between the two
lipid species, and biases the local curvature to be $c_0$ or 0. The upper
length scale cutoff is set by the persistance length $\xi_p \approx d \exp
(4\pi \kappa_c/3k_BT)$, beyond which thermal fluctuations drive the membrane
into a self-avoiding branched polymer phase ($\kappa _c \approx 100 k_B
T_{room}$ for DMPC). In principle $\kappa_c$ depends on the local
concentration $\phi$, but we shall ignore this detail (which allows us to
drop the intrinsic curvature term usually present in Eq.\ \ref{eq:helfrich}).
The energy cost for density fluctuations $\delta \rho$ along the membrane is
${\cal H}_{\rho} = \chi^{-1}_{0} \int (\delta \rho/2)^2 \sqrt g \,d^2 u$
(where $\chi_{0}/\rho^{ 2}_{0}$ is the compressibility of the membrane). 

Our model 2-component membrane \cite{FIRST} consists of a fixed number $N$ of
two types of hard beads A and B (vertices), each of diameter $a$, with $N =
N_A + N_B$, linked together by flexible tethers so as to triangulate a 2-dim
open surface embedded in ${\cal R}^3$. The length of each tether lies between
$a = l_{min} < l < l_{max} = \sqrt 3 a$ (imposes self-avoidance locally
\cite{MC}). The coordination number of every bead is restricted to lie
between 3 and 9. The configurations of our model 2-component membrane are
weighted by the discrete form of ${\cal H}$. The discretised curvature energy
Eq.\ \ref{eq:helfrich} is written as 

\begin{eqnarray}
 &  & \kappa_c \sum_i \sum_{(ij)} \left[ H_{(ij)} +
      \frac {K_i}{\sqrt 3}
      \right]+ {\kappa_0}^2\kappa_c \sum_{i} \frac {(1+\phi_i)^2}
      {4} a_i
      \nonumber \\
 &  & - \kappa_0 \kappa_c \sum_{i} \,(\pm) \sqrt {\sum_{(ij)} \left[
      H_{(ij)} + \frac {K_i}{\sqrt 3}\right]} (1+\phi_i)\,\,,
\label{eq:discrete} 
\end{eqnarray}

 where the index $i$ denotes a vertex,
$(ij)$ the tether connecting $i$ and $j$ and $\sum_{(ij)}$ the sum over all
tethers emanating from $i$. The term $H_{(ij)} = 1-{\hat {\bf n}}_{\alpha}
\cdot {\hat {\bf n}}_{\beta}$\,, where ${\hat {\bf n}}_{\alpha}$ and ${\hat
{\bf n}}_{\beta}$ are the local {\it outward} normals to the triangles
$\alpha$ 

\twocolumn[\hsize\textwidth\columnwidth\hsize\csname @twocolumnfalse\endcsname
\myfigure{\epsfysize5.5in\epsfbox{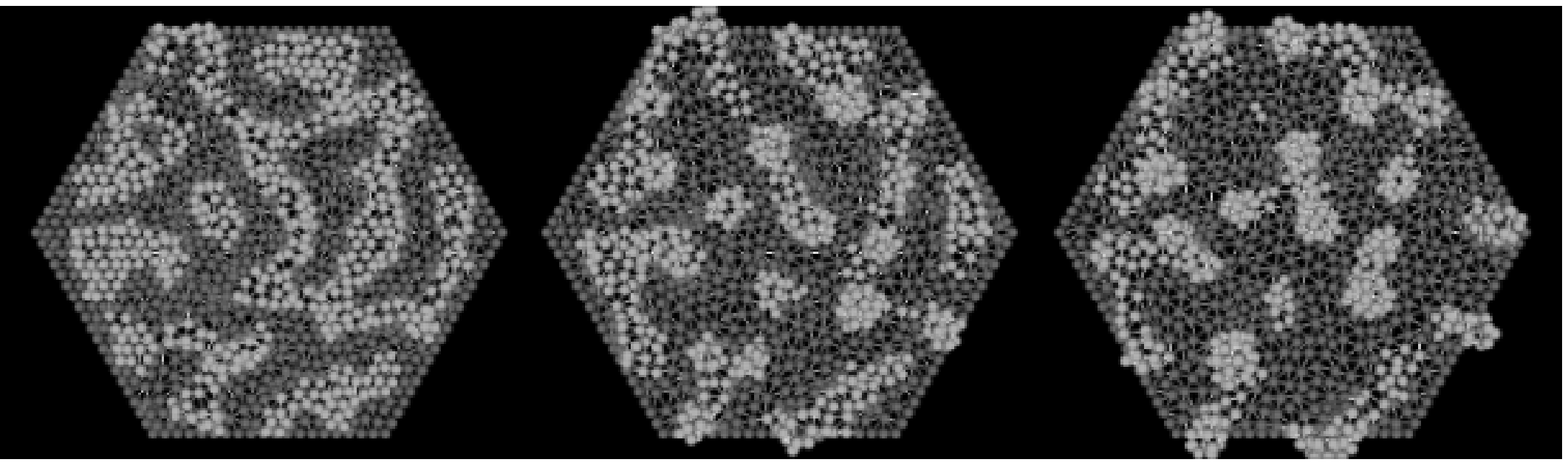}}
{\vskip-3inFig. \ 1~~ Configuration snapshots from the DTMC simulation at
times $t=4000\,,\, 6500\,,\, 17000$ in units of MCS, for a critical quench.
The number of particles is $N=1140$, and the number of flips is $N_{f}=15N$.
The labyrinthine pattern, typical of spinodal decomposition, break up into
buds at late times.}
]

\noindent
and $\beta$ sharing a common tether $(ij)$. The intrinsic curvature
$K_i$, calculated from the deficit angle at $i$, must be added to $H_{(ij)}$
to obtain Eq.\ \ref{eq:helfrich} in the continuum limit. The $\pm$ sign in
the third term, denotes the sign of the local mean curvature --- it is
positive if the {\it outward} normals to adjacent triangles point away from
each other and negative otherwise. The concentration $\phi_i$ takes values
$\pm 1$ depending on whether the vertex $i$ is occupied by A or B. In the
second term of Eq.\ \ref{eq:discrete}, $ a_i$ is the sum of the areas of the
triangles with $i$ as the vertex.  The discrete version of Eq.\
\ref{eq:landau} is clearly, \begin{equation} {\cal H}_{\phi} =
-\,J\,\sum_{<ij>} \phi_{i} \phi_{j}\,\,, \label{eq:ising} \end{equation}
where the sum $<ij>$ is over vertex pairs connected by a tether. Finally,
${\cal H}_{\rho}$ is contained in our imposition of the hard sphere and
tether constraints.

The DTMC for a 2-component fluid membrane consists of bead movements in
${\cal R}^3$ to change the shape \cite{MC}, tether flips to simulate fluidity
(dynamical triangulation) \cite{MC}, and exchange moves between A and B to
promote phase segregation \cite{FIRST}. Moves which violate the hard sphere
and tether constraints, graph planarity (no tether intersections) and
topology of the surface \cite{MC} are rejected. The bead moves and the tether
flips are made according to a Metropolis algorithm with the hamiltonian Eqs.\
\ref{eq:discrete}, \ref{eq:ising}. The exchange moves between A and B are
implemented by the usual Kawasaki dynamics \cite{BRAY}, with a transition
probability $W(i \leftrightarrow j)=[1-\tanh\,(\Delta E/2k_BT)]/2$, where
$\Delta E$ is the energy difference between the final and the initial
configuration. One monte carlo sweep (MCS) is defined as $N$ attempted bead
moves, during which we make $N_{f}$ attempts at flipping tethers and $N_{ex}$
attempts at Kawasaki exchanges. We shall work with two different boundary
conditions --- rigid boundary conditions on a hexagonal frame (fixes the
projected area) and free boundary conditions (unframed), in either case the
tethers along the boundary are not flipped. Details of this algorithm may be
found in \cite{FIRST}. 

A variant of this algorithm was used to study the phase separation dynamics
of dense binary fluids confined to two dimensions \cite{SUNMAD}. The growth
laws and the form of the correlation functions were consistent with a variety
of other simulations \cite{JULIA}. We note that the lateral mobility of the
beads may be tuned by $N_{f}$. 

At time $t=0$, we start with a homogeneous mixture $(N_A=N_B=N/2\,; N=1140)$
of A and B lipids on a flat, randomly triangulated 2-dim surface. After a
quench into the coexistence regime \cite{NOTE2}, the fluid membrane evolves
by the monte carlo dynamics just discussed. We shall investigate the generic
case when the time scales, $\tau$ and $\tau_{\phi}$, over which the shape and
the concentration change are comparable ($N_{ex} = N$), relegating a more
detailed analysis to later \cite{LARGE}. The values of parameters have been
chosen to be $J= \kappa_c = \kappa_0 = 1$ in units of $k_B T = 0.25$. 

Before showing the sequence of monte carlo configurations, let us discuss the
very early time spinodal instability, when we are justified in using a Monge
parametrization of the surface ${\bf R}(u_1,u_2;t) \equiv (x,y,h(x,y;t))$.
The continuum Langevin equations of motion for $h$ and $\phi$ read,
\begin{equation} \frac {\partial \phi}{\partial t} = - c_0 \nabla^4 h -
\sigma \nabla^4 \phi - \nabla^2 \phi + 3 \phi^2 \nabla^2 \phi + 6 \phi
(\nabla \phi)^2 \label{eq:cahn} \end{equation} \begin{equation}
\frac{\tau}{\sqrt{1+(\nabla h)^2}} \frac {\partial h}{\partial t} = -
\kappa_c \nabla^4 h + c_0 \nabla^2 \phi\,\,.  \label{eq:monge} \end{equation}
We perform a linear stability analysis about a flat membrane with $h({\bf
x},t) = \sum_{\bf q} \delta h_{\bf q}(t) e^{i{\bf q}\cdot {\bf x}}$ and
$\phi({\bf x},t) = \phi_0 + \sum_{\bf q} \delta \phi_{\bf q}(t) e^{i{\bf
q}\cdot {\bf x}}$. The overdamped modes are unstable for $q < q^{*}$, where
$q^{*} = \sqrt{(1-3\phi^{2}_0 + c^{2}_0/\kappa_c)/\sigma}$. The small $q$
dispersion of the unstable mode is easily worked out --- $\omega \sim q^{2}$
for $\phi^{2}_0 < 1/3$ and $\omega \sim q^{4}$ for $\phi^{2}_0 > 1/3$. 
Solvent hydrodynamics, introduced through a renormalized $\tau \sim (\eta
q)^{-1}$ ($\eta =$ solvent kinematic viscosity), changes the $q^4$ dispersion
to $\omega \sim q^{5}$, leaving the other features unaffected. The coupling
to shape fluctuations shifts the (mean field) spinodal line closer to the
coexistence line\,: $\phi^{2}_0 = 1/3 + c^{2}_0/3\kappa_c$.  The
configuration coarsens into the usual labyrinthine pattern, Fig.\ 1(a), which
is a snapshot from our DTMC at $t= 4000$ MCS. 

Subsequent evolution depends on the magnitude of the lateral mobility.  Let
us first consider the case when $N_{f} = 15 N$, which corresponds to mobile
lipids in the fluid phase. A look at a time sequence of configurations
generated by our simulation Fig.\ 1 , reveals an unusual scene. The
bicontinuous patches, which are a generic feature of binary fluids at a
critical quench, break up into disconnected buds as time progresses ! Further
coarsening occurs through the motion of these buds, as they approach each
other and subsequently coalesce. The late time behaviour depends on the final
equilibrium configuration of the membrane. 

To understand the breakup of the bicontinuous patches (strip instability), we
again work within the Monge gauge and perform a linear stability analysis
about a flat strip of width $2u_0$ with $\phi=1$ within the strip and
$\phi=-1$ without. We parametrise the perturbed configuration by $h = h_0 +
\sum_{\bf q} \delta h_{\bf q}(t) e^{i(q_1x+q_2y)}$ and $\phi = \tanh
[(u(x,t)-y)/\xi] + \tanh [(u(x,t)+y)/\xi] - 1$, where $u(x,t) = u_0 +
\sum_{q_1} \delta u_{q_1}(t) e^{i{q_1}x}$ and $\xi = \sqrt {2 \sigma}$ is the
interfacial thickness. The time evolution of the fourier amplitudes evaluated
at $y = \pm u_0$ is, to linear order, 

\begin{equation} 
\left[
\begin{array}{c} \delta {\dot u}_{q_1} \\ \delta {\dot h}_{q_1} \end{array}
\right] = \left[ \begin{array}{cc} {-\sigma q^4_1 - 4 q^2_1} &
-\sqrt{2\sigma}c_0 q^4_1 \\ -\frac {c_0}{\tau\sqrt 2\sigma} (\frac
{1}{\sigma} + q^2_1) & -\frac{\kappa_c }{\tau} q^4_1 \end{array} \right]
\left[ \begin{array}{c} \delta {u}_{q_1} \\ \delta {h}_{q_1} \end{array}
\right] \,. \label{eq:matrix} 
\end{equation} 

Such a perturbation leads to a
long wavelength instability of the strip.  The dispersion of the unstable
mode is $\omega \sim q^2_1$, as $q_1\to 0$.  Hydrodynamic coupling to the
solvent via a renormalized $\tau(q)$ does not alter this dispersion. This
instability, arising from a competition between the interfacial and the
curvature energy \cite{JULICHER}, leads to local outward budding since $c_0 >
0$. This linear instability will grow into the complete buds seen in Fig.\
1(c), connected to the rest of the membrane through a narrow neck. The
typical size of the bud is $R_{H} = \kappa_c/(c_0+\sigma)$. The $c_0 = 0$
case is special since the linear analysis does not predict a strip
instability\,; nonlinear terms however would initiate an instability to
budding.  Since the bud could form either outward or inward, the dynamics of
bud formation is logarithmically slow as can be seen by a mapping onto a
1-dim, scalar time-dependent Ginzburg-Landau (TDGL) dynamics \cite{BRAY}. 

\myfigure{\epsfysize2.5in\epsfbox{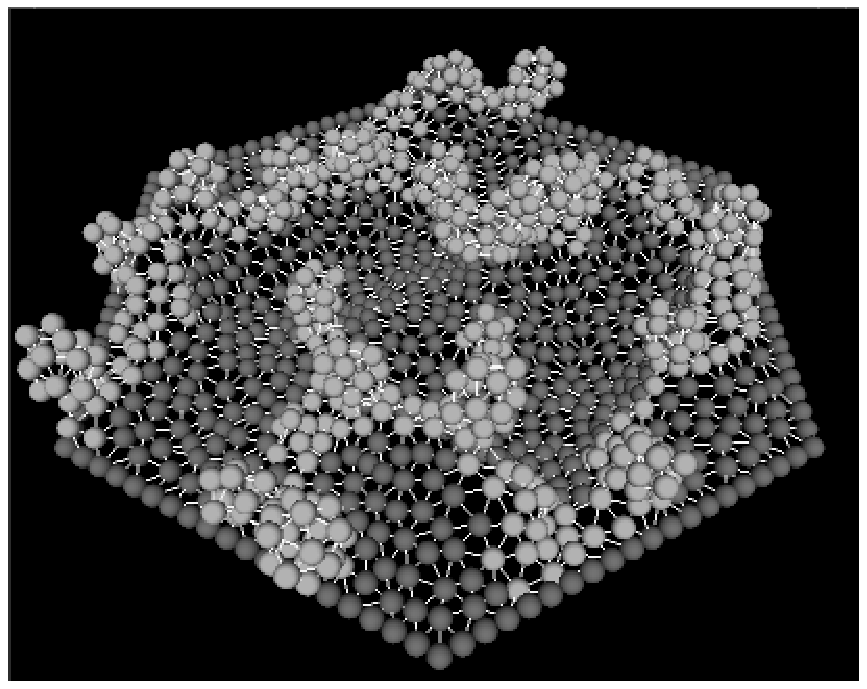}}{\vskip.3inFig. \ 2~~
Configuration snapshot from the DTMC simulation at $t=17000$ MCS for a
critical quench. The number of particles is the same as in Fig.\ 1, but
$N_{f}=N$. The late time configuration is now `crinkled'.} 

When the mobility is small ($N_{f}=N$), the intermediate time shapes are
dramatically different.  Instead of erupting buds, the membrane develops
sharp ridges at the A$\vert$B interface and appears {\it crinkled} (Fig.\ 2)
. We argue that these crinkles are a consequence of a disclination induced
{\it buckling} instability.  The A$\vert$B interfacial energy can be reduced
both by shrinking the interface perimeter and by lowering the local
coordination number of the beads at the interface. This introduces a
preponderance of positive orientational disclination defects at the
interface, in an otherwise 6-fold coordinated network. Though the membrane is
still in the fluid phase \cite{HEXATIC}, the high lipid viscosity $\eta_l$
results in a nonzero shear modulus $\mu \sim \eta_l/\tau_s$ over time scales
smaller than the shear relaxation time $\tau_s$ \cite{LL}. The membrane
therefore responds elastically to the presence of these disclinations.
Placing a positive disclination on a flat elastic sheet of size $R$ costs an
energy proportional to $K_0 R^2$, where $K_0 =
4\mu(\mu+\lambda)/(2\mu+\lambda)$ is related to the elastic moduli.  Buckling
eliminates the elastic contributions to the stress field created by this
defect \cite{BUCKLING}, leaving only a bending energy cost which goes as
$\kappa_c \ln R$. As shown in \cite{BUCKLING}, the elastic membrane buckles
when $K_0 R^2 \geq 160\, \kappa_c$. This physics is consistent with our monte
carlo shape evolution of a flat membrane having a single patch of A in a sea
of B lipids with $N_{f}=N$. The membrane goes into a {\it buckled} conical
shape, with the A species at the tip. This buckled shape is clearly a
long-lived nonequilibrium effect. The same elastic considerations would imply
that negative disclinations would induce a long-lived negatively curved
(saddle) buckled surface. The relative fraction of positive and negative
disclinations depends on the boundary conditions at the frame.  Unless there
is an excess charge, either due to initial conditions or by having an open
boundary, the membrane will be asymptotically flat. The crinkled look of
Fig.\ 2 is a consequence of a random distribution of such positively and
negatively buckled surfaces. 

The dynamics of coarsening and shape changes can be studied from the
behaviour of a variety of two-point correlators. We define as $R_E$, the
length scale associated with the interfacial energy density, $\langle E
\rangle = N^{-1}\sum_{<i,j>} (1-\phi_i \phi_j)$, and $R_H$, the length
extracted from the total mean curvature density $\langle H \rangle = N^{-1}
\sum_{i} \,(\pm) \sqrt {\sum_{(ij)} \left[ H_{(ij)} + \frac {K_i}{\sqrt
3}\right]}$. In addition, we measure $R_{\phi \phi}$ from the first zero of
the intrinsic equal-time correlator for the local concentration
$\Gamma_i(r_2,t) \, \equiv \,\frac{1}{N}\,\sum_{\bf {u}}\,\langle\phi({\bf
u}+{\bf r_2},t)\,\phi({\bf u},t)\rangle$ and $R_{HH}$ from the first zero of
the intrinsic equal-time correlator for the local curvature $\Delta_i(r_2,t)
\, \equiv \,\frac{1}{N}\,\sum_{\bf {u}}\,\langle H({\bf u}+{\bf
r_2},t)\,H({\bf u},t) \rangle$, where $\vert r_2 \vert$ is the geodesic
distance between points ${\bf u}$ and ${\bf u}+{\bf r}_2$ defined on the
2-dim manifold. Geodesic distances are computed using Floyd's algorithm for a
directed graph \cite{ALGOL}. As might be expected from the absence of
self-similarity in Fig.\ 1, we do not observe dynamical scaling in the above
correlators over the time scales of our simulation. 

\myfigure{\epsfysize2.5in\epsfbox{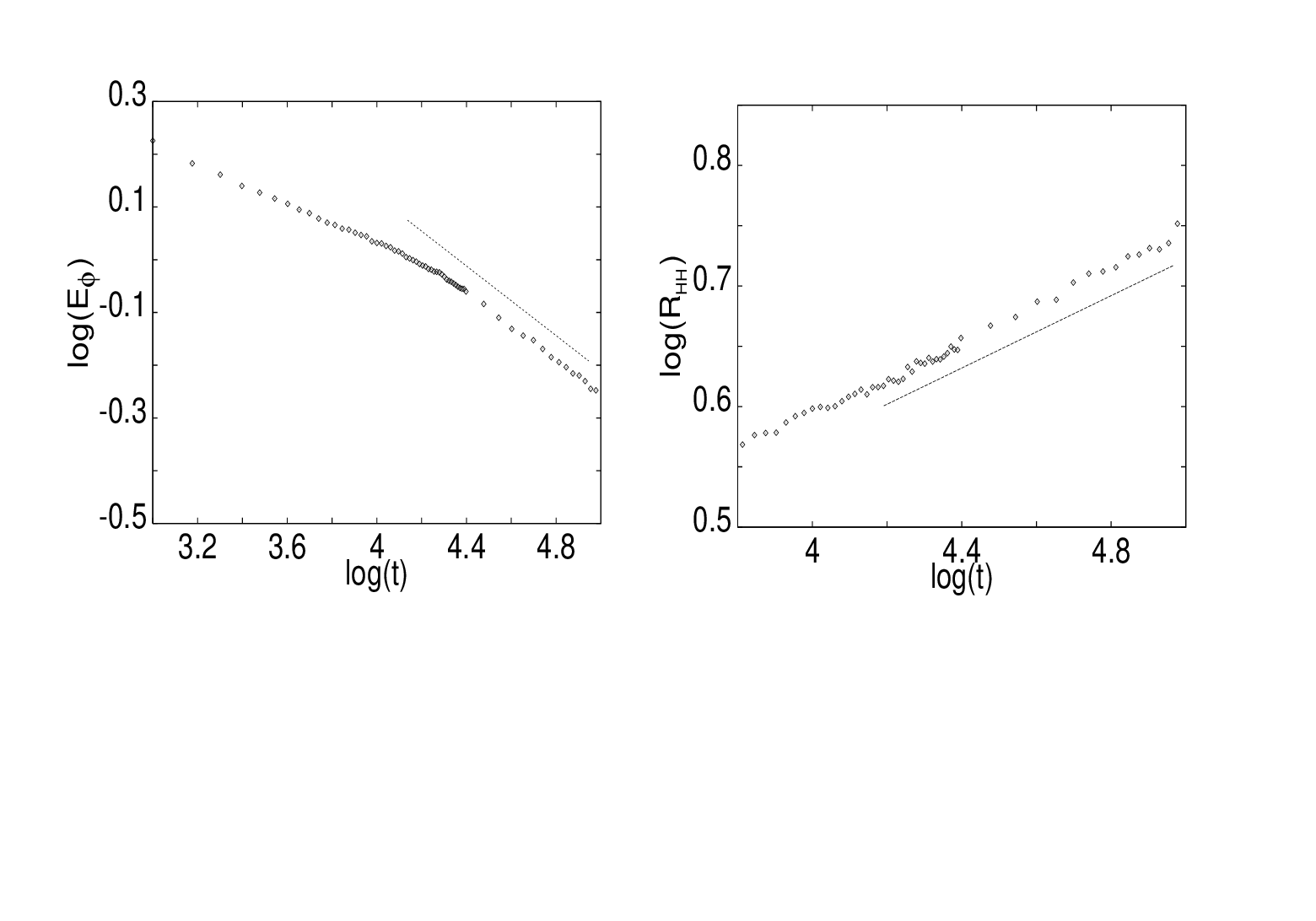}}{\vskip-0.9inFig. \ 3~~ Growth
laws extracted from the DTMC with $N_{f}=15N$ (the straight line fits are
aids to the eye). (a) $R_{E} \sim E_{\phi}^{-1}$ crosses over to the usual
$t^{1/3}$ growth, while (b) $R_{HH}$ exhibits a late time $t^{1/6}$
behaviour.}

The length scales $R_E$, $R_H$ and $R_{HH}$ (averaged over 16 runs) are
displayed in Figs.\ 3(a) and 3(b) for $N_{f} = 15 N$ ($R_{\phi \phi}$ behaves
the same as $R_E$).  We identify two dynamical regimes associated with
distinct growth processes. Regime I heralds the initiation of the strip
instability to the formation of buds and the length scales, $R_E$, $R_H$ and
$R_{HH}$ grow as $t^{1/4}$ \cite{FIRST}. This is followed by regime II where
the buds of typical size $R_H$ move towards each other and subsequently
coalesce. In the absence of any hydrodynamic drag due to the solvent, $R_E
\sim t^{1/3}$. This dependence will be unchanged by inclusion of a viscous
drag due to the solvent (though the prefactor would be altered). The buds
carry both $\phi$ and curvature $H$, and so equating the energy densities
$E_{\phi} \sim R^{-1}_{E}$ and $E_{H} \sim R^{-2}_{HH}$, leads to $R_{HH}
\sim R^{1/2}_{E} \sim t^{1/6}$ (Fig.\ 3(b)). 

It would be interesting to carry out a systematic experimental study of
dynamical morphology changes and shape instabilities in mixed artificial
membranes. Ofcourse the choice of the lipid species is crucial\,; it is
important to ensure that $T_c > T_m$, the main transition temperature below
which the lipid freezes into a gel state ($T_m \approx 23^{\circ}C$ for
DMPC). This may be achieved by choosing lipids with short chains or with a
lot of unsaturated bonds within the chain \cite{CELL}.  The dynamical
buckling phenomenon reported above could be observed by inducing
photochemical (UV) crosslinking of the minority lipids after they have
clustered. The present study might be of relevance to biological membranes as
well, in particular to the dynamics of `budding' of coated vesicles and the
`patching' of membrane proteins and glycolipids \cite{CELL}. 
 

\begin{references}
\bibitem{NOTE2} The critical temperature $T_c \to 0$ in the thermodynamic
limit, since for sizes $L \gg \xi_p$, the membrane behaves like a
self-avoiding branched polymer, with the lipids sitting on the boundary. For
DMPC membranes, however, $\xi_p$ is fairly large.
\bibitem{HEXATIC} The presence of a number of isolated disclinations, and a
structureless $\langle n_5(r)n_7(0)\rangle$ ($n_5$ and $n_7$ are the
densities of the 5- and 7- membered rings respectively), imply that the
membrane is not in the hexatic phase.
\end{references}
\end{document}